\documentclass[aps,pra,twocolumn,superscriptaddress,10pt,showkeys,fleqn]{revtex4-2}

\usepackage{amsmath,amsfonts,amssymb,amsthm}
\usepackage{graphicx}
\usepackage[colorlinks=true,urlcolor=blue,citecolor=blue,linkcolor=blue]{hyperref}
\usepackage{bbm}

\begin{document}

\title{Entangled and disentangled multi-time quantum trajectories in relativistic systems}

\author{Yannick Noel Freitag}
\email{yannick.freitag@upb.de}
\affiliation{
    Theoretical Quantum Science, Institute for Photonic Quantum Systems (PhoQS), Paderborn University, Warburger Stra\ss{}e 100, 33098 Paderborn, Germany
}

\author{Julien Pinske}
\affiliation{
    Niels Bohr Institute, University of Copenhagen, Jagtvej 155 A, DK-2200 Copenhagen, Denmark
}

\author{Jan Sperling}
\affiliation{
    Theoretical Quantum Science, Institute for Photonic Quantum Systems (PhoQS), Paderborn University, Warburger Stra\ss{}e 100, 33098 Paderborn, Germany
}

\date{\today}

\begin{abstract}
    Quantum entanglement is a key resource for quantum technologies, including emerging ground-to-satellite quantum communication.
    In such a scenario, an important challenge to be overcome is to consider entanglement between two or more quantum particles in different inertial frames, potentially experiencing relativistic effects affecting quantum correlations.
    In this paper, we present a consistent framework that overcomes this challenge.
    To this end, we establish the notion of factorizable and entangled multi-time trajectories and derive a class of Euler--Lagrange equations under the constraint of a non-entangling behavior.
    Comparing this restricted evolution to the solutions of the unrestricted equations of motion allows us to investigate the trajectory-based entanglement of general systems.
    We solve our equations for interacting particles in a Klein--Gordon-type setting, thereby quantifying the dynamic and relativistic impact of entanglement in a self-consistent manner.
\end{abstract}

\maketitle

%%%%%%%%%%%%%%%%%%%%%%%%%%%%%%%%%%%%%%%%%%%%%%%%%%%%%%%%%%%
%%%%%%%%%%%%%%%%%%%%%%%%%%%%%%%%%%%%%%%%%%%%%%%%%%%%%%%%%%%
%%%%%%%%%%%%%%%%%%%%%%%%%%%%%%%%%%%%%%%%%%%%%%%%%%%%%%%%%%%

\section{Introduction}
\label{sec:Introduction}

    Today's strong interest in the fundamental notion of quantum entanglement can be attributed to its essential role in quantum information processing and quantum communication \cite{HH09, NC10}.
    For instance, entanglement is beneficial for pioneering applications, including, e.g., quantum computing, quantum cryptography, quantum teleportation \cite{GRTZ02, BPMEWZ97, BBCJPW93}.
    Furthermore, entanglement is the foundation for future large-scale quantum networks \cite{KW19}, for example, utilizing quantum repeaters \cite{Letal19, KRHBKNBTMH17} and atmospheric channels \cite{VSV16}.
    To realize large-distance, earth-bound quantum communication, however, quantum repeaters would need to be available in large numbers while noise and turbulence induced by the atmosphere render transmissions upwards of hundred kilometers difficult.
    One realization that avoids these issues is ground-to-satellite quantum communication, as recently demonstrated in experiments \cite{Letal17, Retal17}, because the thickness of the atmosphere is only in the order of tens of kilometers.

\begin{figure}[b]
    \includegraphics[width=.5\columnwidth]{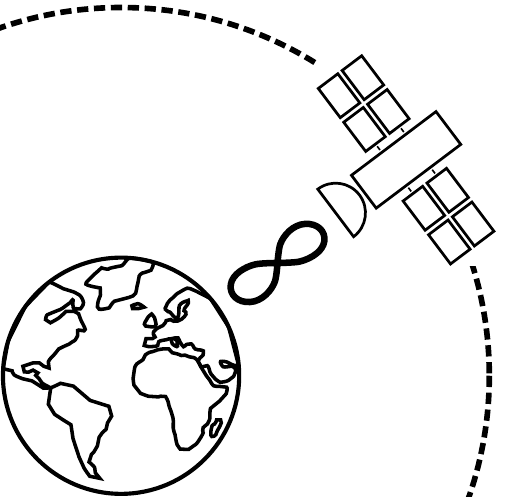}
    \caption{%
        Ground-to-satellite quantum communication with sender and receiver in distinct non-inertial systems.
        Entanglement may be subject to effects caused by non-comoving reference frames.
    }\label{fig:scheme}
\end{figure}

    Relativistic effects become relevant in ground-to-satellite quantum links, which are well-explored in the non-quantum context of the global positioning systems \cite{P-S07}, for instance.
    Such effects include time dilation, caused by the high speed of satellites and the Earth's gravitational field, as well as rotational phenomena, like the Sagnac effect.
    Quantum dynamics and quantum correlations in such non-trivial reference frames present a current obstacle;
    see Fig. \ref{fig:scheme}.
    In fact, it has been proposed that quantum optical experiments with satellites can be used to probe the boundary between quantum mechanics and relativity \cite{Retal12}.

    The time-dependent description of quantum systems, including relativistic effects, is one subject of contemporary research.
    Studying quantum phenomena in non-inertial frames led to the discovery of both the Unruh effect \cite{U76} and Hawking radiation \cite{H74} and has inspired further studies, highlighting the interplay of general relativity and quantum mechanics \cite{FMM-MM10,BLM-MDF10}.
    Also, in Refs. \cite{PT04,MR12,FLTSSJF13}, the importance of relativistic effects for quantum information was emphasized.
    Moreover, it was suggested that entanglement in relativistic scenarios can be observer-dependent \cite{AF12}.
    In addition, non-inertial frames can have meaningful, observable impacts, such as a delay in the well-known Hong--Ou--Mandel interference \cite{RTGUFP19}, and can inspire other applications, e.g., a proposal for a quantum version of the Cavendish experiment \cite{BCM18}.
    For practical applications, studies show that entanglement distillation protocols benefit from time-dependent approaches to the entanglement problem \cite{LM-DS06}.

    The need for a consistent framework to investigate entanglement of spacetime trajectories also becomes apparent when considering certain, seemingly contradicting findings in the literature.
    Some works show that the degree of violation of Bell's inequality is decreasing with increasing velocity of the observer \cite{ALMH03}.
    On the other hand, it was claimed that entanglement is a Lorentz-invariant quantity \cite{Y10, F10}.
    Other work shows that entanglement can even remain maximal for all moving observers and inertial frames \cite{LD03,BBENFBFHU24}.
    New concepts, like quantum reference frames, have been developed, in which entanglement becomes a frame-dependent feature \cite{GC-RB19}.

    In this paper, we overcome aforementioned limitations by constructing a consistent framework to describe relativistic entanglement.
    To this end, a frame-independent notion of factorizable quantum trajectories is introduced, which exceeds the notion of factorizable states.
    For this concept, we then derive equations of motion, dubbed relativistic factorizability Euler--Lagrange (ReFEL) equations, that force the particles' trajectories to remain factorizable under arbitrary spacetime transformations.
    To demonstrate the method's capabilities, we study the buildup of entanglement due to interaction between relativistic particles in a proof-of-concept model.
    The comparison of both the restricted and unrestricted evolution yields insight into the entangling dynamics.

    The paper is organized as follows.
    Section \ref{sec:Preliminaries} provides a brief overview of the techniques used, also motivating the need for a consistent framework through an instructive counterexample.
    In Sec. \ref{sec:Novelties}, we derive our ReFEL equations, unambiguously identifying entanglement in arbitrary frames.
    The methodology is applied by comparing the factorizable and entangled quantum trajectories for interacting particles in a Klein--Gordon-like setting in Sec. \ref{sec:KleinGordon}.
    We conclude in Sec. \ref{sec:Conclusion}.

%%%%%%%%%%%%%%%%%%%%%%%%%%%%%%%%%%%%%%%%%%%%%%%%%%%%%%%%%%%
%%%%%%%%%%%%%%%%%%%%%%%%%%%%%%%%%%%%%%%%%%%%%%%%%%%%%%%%%%%
%%%%%%%%%%%%%%%%%%%%%%%%%%%%%%%%%%%%%%%%%%%%%%%%%%%%%%%%%%%
\section{Factorizable and entangled, multi-time trajectories}
\label{sec:Preliminaries}

\subsection{Definition}

    In a time-independent picture, the Dirac--von~Neumann axioms state that a system is described by a vector in Hilbert space, $|\psi\rangle\in\mathcal H$.
    From the same set of axioms, we have $\mathcal H=\mathcal H_1\otimes\cdots\otimes\mathcal H_N$ for a composite system of $N$ particles.
    A factorizable state reads
    \begin{equation}
        \label{eq:FacStates}
        |\psi_\mathrm{fac}\rangle
        =
        |\psi_1\rangle\otimes\cdots\otimes|\psi_N\rangle,
    \end{equation}
    and $|\psi\rangle\neq|\psi_\mathrm{fac}\rangle$ defines entanglement.
    This time-independent and state-based concept is the common scenario in which one studies entanglement \cite{NC10, BL06, B09}.

    Introducing time $\bar t$, we can describe the collection of states over time via the trajectory $
        \left\{
            |\psi(\bar t)\rangle
        \right\}_{
            \bar t\in\mathbb R
        }
    $, using a notation as a continuous list of Hilbert space vectors that is adapted from the analogous notion for stochastic processes.
    A trajectory of a product state reads
    \begin{equation}
        \label{eq:FacTrajectory}
        \left\{
            |\psi_1(\bar t)\rangle\otimes\cdots\otimes|\psi_N(\bar t)\rangle
        \right\}_{
            \bar t\in\mathbb R
        }.
    \end{equation}

    In 1932, Dirac \cite{D32} proposed a multi-time approach to quantum mechanics, in which each particle is described with its own time coordinate \cite{LPT17,LPT20}.
    Then, we have a multi-time state $|\psi(t_1,\ldots,t_N)\rangle$ which may be related to the commonly-used single-time scenario by setting all times to a single one, $|\psi(\bar t,\ldots,\bar t)\rangle=|\psi(\bar t)\rangle$.
    With $|\psi_\mathrm{fac}(t_1,\ldots,t_N)\rangle=|\psi_1(t_1)\rangle\otimes\cdots\otimes|\psi_N(t_N)\rangle$, we can extend the factorization of states to the factorization of the trajectories as follows:
    \begin{equation}
        \label{eq:MultiTimeFacTrajectory}
    \begin{aligned}
        &\left\{
            |\psi_\mathrm{fac}(t_1,\ldots,t_N)\rangle
        \right\}_{
            t_1,\ldots,t_N\in\mathbb R
        }
        \\
        ={}&
        \left\{ |\psi_1(t_1)\rangle \right\}_{ t_1\in\mathbb R }
        \otimes\cdots\otimes
        \left\{ |\psi_N(t_N)\rangle \right\}_{ t_N\in\mathbb R }.
    \end{aligned}
    \end{equation}

    For a single time, $t_1=\cdots=t_N=\bar t$, the factorized trajectories in Eq. \eqref{eq:MultiTimeFacTrajectory} take the form of a trajectory of a product state in Eq. \eqref{eq:FacTrajectory}.
    Otherwise, the factorization of states in Eq. \eqref{eq:FacStates} can be elevated to the factorization of trajectories in Eq. \eqref{eq:MultiTimeFacTrajectory}.
    In other words, Dirac's multi-time approach renders it possible to define entanglement of trajectories in an analogous manner as done for time-independent states,
    \begin{equation}
        \label{eq:OurEntanglement}
    \begin{aligned}
        &
        \left\{
            |\psi(t_1,\ldots,t_N)\rangle
        \right\}_{
            t_1,\ldots,t_N\in\mathbb R
        }
        \\
        \neq{}&
        \left\{ |\psi_1(t_1)\rangle \right\}_{ t_1\in\mathbb R }
        \otimes\cdots\otimes
        \left\{ |\psi_N(t_N)\rangle \right\}_{ t_N\in\mathbb R }.
    \end{aligned}
    \end{equation}
    An entangled trajectory now encompasses both:
    we may fail to factorize the state vector in Hilbert space, and we may fail to separate trajectories in time coordinates.

    We remark that, in the context of mixed states, entanglement is superseded by the notion of inseparable states \cite{W89}.
    Throughout this work, however, we focus on pure states.
    Still, a generalization can be done via convex combinations of pure states, such as argued in Ref. \cite{SW20}.
    As a preliminary summary, the approach considered here is based on a tensor product of trajectories, cf. Eq. \eqref{eq:MultiTimeFacTrajectory}, rather than a trajectory of tensor products, cf. Eq. \eqref{eq:FacTrajectory}.

\subsection{Example}

    To emphasize why the factorization of trajectories supersedes the separation of states, we consider the two-time, two-particle wave function $\psi(t_1,x_1,t_2,x_2)=f_1(t_2,x_1)f_2(t_1,x_2)$;
    mind the swapped time indices in the arguments of the two functions $f_1$ and $f_2$.
    Clearly, for all time coordinates $t_1$ and $t_2$, the positions of the particles factorize;
    i.e., $|\psi_j(t_{3-j})\rangle=\int \mathrm{d}x_j f_j(t_{3-j},x_j)|x_j\rangle$ for $j\in\{1,2\}$ describes the local state of particle $j$,
    \begin{equation}
        \label{eq:Counterexample}
        |\psi(t_1,t_2)\rangle=|\psi_1(t_2)\rangle\otimes|\psi_2(t_1)\rangle,
    \end{equation}
    thus showing no spatial entanglement of the states.
    But the trajectories are entangled as the time of the second particle describes the temporal behavior of the first particle.
    In terms of trajectories, we can express this temporal entanglement as
    \begin{equation}
    \begin{aligned}
        &\left\{|\psi_1(t_2)\rangle\right\}_{t_2\in\mathbb R}
        \otimes
        \left\{|\psi_2(t_1)\rangle\right\}_{t_1\in\mathbb R}
        \\
        \neq{}&
        \left\{|\psi_1(t_1)\rangle\right\}_{t_1\in\mathbb R}
        \otimes
        \left\{|\psi_2(t_2)\rangle\right\}_{t_2\in\mathbb R}.
    \end{aligned}
    \end{equation}
    Again mind the swapped time indices here.
    Thus, the joint trajectory for the example under study is a trajectory of products but not a product trajectory.

    The deviation of the constructed state from our definition of factorized trajectories in Eq. \eqref{eq:MultiTimeFacTrajectory} leads to interesting consequences in a relativistic setting.
    On the one hand, the joint state is factorizable, in the sense of Eq. \eqref{eq:FacStates}, for all times, $\psi(t_1,x_1,t_2,x_2)=f_1(t_2,x_1)f_2(t_1,x_2)$, as discussed above.
    On the other hand, we can apply Lorentz transformations to both coordinate systems, even identical ones,
    \begin{equation}
    	x_j \mapsto \frac{x_j + v t_j}{\sqrt{1-v^2}},
    	\quad
    	t_j \mapsto \frac{t_j + v x_j}{\sqrt{1-v^2}},
    \end{equation}
    with the velocity $v$ and the speed of light set to one, $c=1$.
    Already for the simple case $t_1=t_2=0$, this yields a transformed wave function
    \begin{equation}
    \begin{aligned}
        \psi(t_1,x_1,t_2,x_2)
        \mapsto{}&
        f_1\left(\frac{vx_2}{\sqrt{1-v^2}},\frac{x_1}{\sqrt{1-v^2}}\right)
        \\
        {}&\times
        f_2\left(\frac{vx_1}{\sqrt{1-v^2}},\frac{x_2}{\sqrt{1-v^2}}\right),
    \end{aligned}
    \end{equation}
    which is, in general, not a product $\psi_1(x_1)\psi_2(x_2)$.
    As an illustrative case, consider specifying the joint state introduced above as a Gaussian $\psi(t_1,x_1,t_2,x_2)=\exp(-(x_1-t_2)^2)\exp(-(x_2-t_1)^2)$.
    After applying the Lorentz transformations, and setting $t_1=t_2=0$, the transformed function becomes
    \begin{equation}
    \begin{aligned}
        \psi(t_1,x_1,t_2,x_2)
        \mapsto{}&
        \exp\left(-\frac{(x_1-vx_2)^2}{1-v^2}\right)
        \\
        {}&\times
        \exp\left(-\frac{(x_2-vx_1)^2}{1-v^2}\right),
    \end{aligned}
    \end{equation}
    which is clearly not a product state since the last term in $\exp(-\frac{1+v^2}{1-v^2}x_1^2)\exp(-\frac{1+v^2}{1-v^2}x_2^2)\exp(\frac{4v}{1-v^2}x_1x_2)$ is not factorizable.

    Therefore, even a joint change of reference frames for both particles takes the spatially factorizable state and makes it spatially entangled.
    In our context of trajectories, the explanation is straightforward:
    the temporal entanglement of the trajectories is mapped to the spatial one with the boost, not altering the overall spacetime entanglement properties.
    Without the notion of temporal quantum correlations in the trajectory, however, the sudden appearance of spatial entanglement when the particles jointly undergo a Lorentz transformation would be rather surprising.

\subsection{Discussion}

    The resolution of ambiguous factorizations of states through factorizations of trajectories is exactly what our framework aims to refine.
    The attribute of factorizability or entanglement should be invariant under any spacetime transformation acting on individual trajectories.
    This is analogous to local unitary transformations of states in standard (i.e., time-independent) entanglement theory.
    Here, for every $j$, any Hilbert-spacetime transformation $\Lambda_j$ that acts on individual particles,
    \begin{equation}
        \Lambda_j: \quad (t_j,|\psi_j(t_j)\rangle)\mapsto (t'_j,|\psi'_j(t'_j)\rangle),
    \end{equation}
    cannot change the factorizability of trajectories as defined in Eq. \eqref{eq:MultiTimeFacTrajectory}, i.e.,
    $
        \left\{|\psi_1(t_1)\rangle\right\}_{t_1\in\mathbb R}
        \otimes
        \left\{|\psi_2(t_2)\rangle\right\}_{t_2\in\mathbb R}
        \mapsto
        \left\{|\psi'_1(t'_1)\rangle\right\}_{t'_1\in\mathbb R}
        \otimes
        \left\{|\psi'_2(t'_2)\rangle\right\}_{t'_2\in\mathbb R}
    $.
    This spacetime-transformation invariance is not specific to the aforementioned example.
    For all product maps of independent/spacetime-local transformations $\Lambda_1 \otimes\cdots\otimes\Lambda_N$ the product structure of trajectories is preserved by construction.
    They may alter the spatial entanglement at fixed times, as we have seen in the example above,
    but cannot alter the overall entanglement of the spacetime trajectory.
    This also includes maps that take particles to non-inertial frames.

    For the above example, the wavefunction chosen in Eq. \eqref{eq:Counterexample} is spatially factorizable, whilst its trajectory is entangled.
    This temporal entanglement is mapped to a spatial entanglement via Lorentz transformations on individual particles.
    The attribute of spatial entanglement depends upon the frame of reference, which we deem undesirable.
    Our notion of trajectory entanglement always identifies the studied trajectory as spacetime- or trajectory-entangled, regardless of the frame of reference.
    Note that, while not considered explicitly, our approach also applies for spinor-valued and vector-valued wave functions.

    After introducing the notion of factorizable trajectories, equations that determine the factor trajectories $\{|\psi_j(t_j)\rangle\}_{t_j\in\mathbb R}$ from initial states are needed.
    This is achieved with the approach in the next section.

%%%%%%%%%%%%%%%%%%%%%%%%%%%%%%%%%%%%%%%%%%%%%%%%%%%%%%%%%%%
%%%%%%%%%%%%%%%%%%%%%%%%%%%%%%%%%%%%%%%%%%%%%%%%%%%%%%%%%%%
%%%%%%%%%%%%%%%%%%%%%%%%%%%%%%%%%%%%%%%%%%%%%%%%%%%%%%%%%%%
\section{Equations of motion}
\label{sec:Novelties}

\subsection{Derivation}

    Recall that analytical mechanics allows us to determine trajectories from the variation of the action $
        S=\int \mathrm{d}t_1\cdots\mathrm{d}t_N\,L
    $, where $L$ is the Lagrangian which depends on the ket-state vector $|\psi(t_1,\ldots,t_N)\rangle$, its derivatives $\partial_{t_j} |\psi\rangle$, and their bra-vector counterparts.
    The underlying principle of stationary action then yields the Euler--Lagrange equations for $N$ times variables as
    \begin{equation}
        \label{eq:ELE}
        0=
        \frac{\partial L}{\partial \langle\psi|}
        -\partial_{t_1}
        \frac{\partial L}{\partial(\partial_{t_1} \langle\psi|)}
        -\cdots
        -\partial_{t_N}
        \frac{\partial L}{\partial(\partial_{t_N} \langle\psi|)}.
    \end{equation}

    Now suppose $|\psi_\mathrm{fac}\rangle=|\psi_1(t_1)\rangle\otimes\cdots\otimes|\psi_N(t_N)\rangle$ as one multi-time element of the factorizable trajectory in Eq. \eqref{eq:MultiTimeFacTrajectory}.
    The action can be recast into the form
    $
        S=\int \mathrm{d}t_j\,L_j
    $,
    where the effective Lagrangian for the subsystem $j$ reads
    \begin{equation}
        \label{eq:EffectiveLagrangian}
        L_j
        =
        \int
        \mathrm{d}t_1\cdots \mathrm{d}t_{j-1}
        \mathrm{d}t_{j+1}\cdots \mathrm{d}t_N\,L.
    \end{equation}
    The subsystem Lagrangian explicitly depends on $|\psi_j(t_j)\rangle$ while the other components have been integrated over, i.e., traced out.
    For the remaining, single-time state $|\psi_j(t_j)\rangle$, the variation to obtain a stationary action yields
    \begin{equation}
        \label{eq:ReFEL}
        0 =
        \frac{\partial L_j}{\partial \langle\psi_j|}
        -\partial_{t_j} \frac{\partial L_j}{\partial(\partial_{t_j}\langle\psi_j|)}
    \end{equation}
    for each $j\in\{1,\ldots,N\}$.

    The equation of motion \eqref{eq:ELE} for the unconstrained trajectory can be compared with the $N$ equations of motion \eqref{eq:ReFEL} that are constrained to factorizable trajectories.
    The $N$ constraint equations collectively describe the multi-time factors $|\psi_1(t_1)\rangle\otimes\cdots\otimes|\psi_N(t_N)\rangle$ and result in a factorizable trajectory.
    How much the unconstrained solution deviates from the factorizable one then determines the trajectories entanglement.

    Further note that, for $t_1=\cdots=t_N=\bar t$, Eq. \eqref{eq:ReFEL} is identical with the approach from Ref. \cite{SW17} that is limited to spatial factorizability.
    Furthermore, in a closed system, a technique from Ref. \cite{SW20} allows one to extend our equations of motion to an ensemble of trajectories of mixed states.
    It is also noteworthy that our approach has loose resemblance with the Hartree--Fock method for many-body systems.
    Therein, one describes the dynamics of a single particle in terms of an effective potential that incorporates interactions with other particles.

\subsection{Integral representation}

    Utilizing the wave function representation of each factor, $|\psi_j(t_j)\rangle=\int\mathrm{d}x_j\psi_j(t_j,x_j)|x_j\rangle$, the Lagrangian is expressed via the Lagrangian density, $L=\int\mathrm{d}x_1\cdots\mathrm{d}x_N\,\mathcal L$.
    Equation \eqref{eq:ReFEL} for spacetime coordinates takes the form
    \begin{equation}
        \label{eq:densityReFEL}
        0
        =\frac{\partial\mathcal L_j}{\partial\psi_j^\ast}
        -\partial_{t_j}\frac{\partial\mathcal L_j}{\partial(\partial_{t_j}\psi_j^\ast)}
        -\partial_{x_j}\frac{\partial\mathcal L_j}{\partial(\partial_{x_j}\psi_j^\ast)},
    \end{equation}
    where $\mathcal L_j$ is obtained by integrating $\mathcal L$ over all but the $j$th spacetime coordinates, similarly to Eq. \eqref{eq:EffectiveLagrangian}.

    The goal is to rewrite Eq. \eqref{eq:densityReFEL} in terms of subsystem projections.
    To this end, notice the following relations for $\psi_\mathrm{fac}=\psi_1\cdots\psi_N$ and the original Lagrange density $\mathcal L$ via the chain rule:
    \begin{equation}
    \begin{aligned}
        \frac{\partial \mathcal L}{\partial \psi_j^\ast}
        ={}&
        \frac{\partial \psi_\mathrm{fac}^\ast}{\partial \psi_j^\ast}
        \frac{\partial \mathcal L}{\partial \psi_\mathrm{fac}^\ast}
        +\sum_{k}
            \frac{\partial (\partial_{t_k}\psi_\mathrm{fac}^\ast)}{\partial \psi_j^\ast}
            \frac{\partial \mathcal L}{\partial(\partial_{t_k} \psi_\mathrm{fac}^\ast)}
        \\
        {}&
        +\sum_{k}
            \frac{\partial (\partial_{x_k}\psi_\mathrm{fac}^\ast)}{\partial \psi_j^\ast}
            \frac{\partial \mathcal L}{\partial(\partial_{x_k} \psi_\mathrm{fac}^\ast)}
        \\
        ={}&
        \frac{\psi_\mathrm{fac}^\ast}{\psi_j^\ast}\frac{\partial\mathcal L}{\partial\psi_\mathrm{fac}^\ast}
        +\sum_{k:k\neq j}
            \frac{\partial_{t_k}\psi_\mathrm{fac}^\ast}{\psi_j^\ast}
            \frac{\partial \mathcal L}{\partial(\partial_{t_k} \psi_\mathrm{fac}^\ast)}
        \\
        {}&
        +\sum_{k:k\neq j}
            \frac{\partial_{x_k}\psi_\mathrm{fac}^\ast}{\psi_j^\ast}
            \frac{\partial \mathcal L}{\partial(\partial_{x_k} \psi_\mathrm{fac}^\ast)}
    \end{aligned}
    \end{equation}
    and
    \begin{equation}
    \begin{aligned}
        \frac{\partial\mathcal L}{\partial(\partial_{q_j}\psi_j^\ast)}
        ={}&
        \frac{\partial(\partial_{q_j}\psi_\mathrm{fac}^\ast)}{\partial(\partial_{q_j}\psi_j^\ast)}
        \frac{\partial\mathcal L}{\partial(\partial_{q_j}\psi_\mathrm{fac}^\ast)}
        \\
        ={}&\frac{\psi_\mathrm{fac}^\ast}{\psi_j^\ast}
        \frac{\partial\mathcal L}{\partial(\partial_{q_j}\psi_\mathrm{fac}^\ast)},
    \end{aligned}
    \end{equation}
    with $q\in\{t,x\}$ and $\psi_\mathrm{fac}/\psi_j=\psi_1\cdots\psi_{j-1}\psi_{j+1}\cdots\psi_N$.
    Integrating over all spacetime coordinates but the $j$th one and applying integration by parts, $\int \mathrm{d}q_k\,(\partial_{q_k}\psi_k^\ast)f=\psi_k^\ast f-\int \mathrm{d}q_k\,\psi_k^\ast (\partial_{q_k}f)$, with vanishing boundary terms, we find
    \begin{equation}
        \label{eq:ReFEL2}
    \begin{aligned}
        0={}&
        \int\mathrm{d}t_1\mathrm{d}x_1\,\psi_1^\ast
        \cdots
        \int\mathrm{d}t_{j-1}\mathrm{d}x_{j-1}\,\psi_{j-1}^\ast
        \\
        {}&\times
        \int\mathrm{d}t_{j+1}\mathrm{d}x_{j+1}\,\psi_{j+1}^\ast
        \cdots
        \int\mathrm{d}t_N\mathrm{d}x_N\,\psi_N^\ast
        \\
        {}&
        \times\left(\frac{\partial\mathcal{L}}{\partial\psi_\mathrm{fac}^\ast}
        -\partial_{t_1}\frac{\partial\mathcal{L}}{\partial(\partial_{t_1} \psi_\mathrm{fac}^\ast)}
        \cdots
        -\partial_{t_N}\frac{\partial\mathcal{L}}{\partial(\partial_{t_N} \psi_\mathrm{fac}^\ast)}
        \right.
        \\
        {}&
        \left.
        -\partial_{x_1}\frac{\partial\mathcal{L}}{\partial(\partial_{x_1} \psi_\mathrm{fac}^\ast)}
        \cdots
        -\partial_{x_N}\frac{\partial\mathcal{L}}{\partial(\partial_{x_N} \psi_\mathrm{fac}^\ast)}
        \right).
    \end{aligned}
    \end{equation}

    Equation \eqref{eq:ReFEL2} is a rephrasing of Eq. \eqref{eq:densityReFEL} and contains an integrand akin to the one in the original Euler--Lagrange equation for $\psi=\psi_\mathrm{fac}$
    [cf. Eq. \eqref{eq:ELE} for Lagrange densities].
    In addition, however, Eq. \eqref{eq:densityReFEL} includes integrals of the form $\int \mathrm{d}t_k\mathrm{d}x_k\, \psi_k^\ast f$ for $k\neq j$ that correspond to projections onto trajectory factors, $\{\langle\psi_k(t_k)|\}_{t_k\in\mathbb R}$.
    For a given Lagrangian density $\mathcal L$, Eq. \eqref{eq:ReFEL2} can be solved and compared to the one without projections, i.e., without factorizability constraints, to probe spacetime entanglement of the quantum trajectory.

    With the here-derived equations of motion, we can determine the factors $\{|\psi_j(t_j)\rangle\}_{t_j\in\mathbb R}$ for our notion of factorizable trajectories.
    Next, we apply both approaches to a second-order Klein--Gordon example, while also constructing trajectory-entanglement witnesses.
    We emphasize that the general derivation of the ReFEL equations does not rely on the specific choice of the Lagrangian, and it applies to any dynamics that admits a suitable Lagrangian formulation.
    For example, for Dirac particles, one would instead start from the first-order bispinor Lagrangian and apply the same procedure as demonstrated in the  above example.
    The resulting ReFEL equations are first-order and spinor-valued equations of motion for each particle.
    Importantly, the preservation of trajectory factorizability by local spacetime transformations---including its spinor representation---remains unchanged.

%%%%%%%%%%%%%%%%%%%%%%%%%%%%%%%%%%%%%%%%%%%%%%%%%%%%%%%%%%%
%%%%%%%%%%%%%%%%%%%%%%%%%%%%%%%%%%%%%%%%%%%%%%%%%%%%%%%%%%%
%%%%%%%%%%%%%%%%%%%%%%%%%%%%%%%%%%%%%%%%%%%%%%%%%%%%%%%%%%%
\section{Application: Interacting Klein--Gordon particles}
\label{sec:KleinGordon}

\subsection{Proof-of-concept model}
\label{sec:Model}

    For benchmarking our methodology, we here apply Eq. \eqref{eq:ReFEL2} to a model of interacting particles obeying a Klein--Gordon-like equation.
    As interactions commonly produce entanglement, the solutions of the conventional Euler--Lagrange equations generally differ from the constrained ones.
    By comparison of both cases, the entanglement of trajectories can then be studied.

    In a covariant description, the Lagrangian density under study reads
    \begin{equation}
        \label{eq:ExtKG}
        \mathcal{L} 
        = \sum_{k=1}^N
        \left(
            g^{\alpha_k\beta_k}(\partial_{\alpha_k}\psi^\ast)(\partial_{\beta_k}\psi)-m_k^2|\psi|^2
        \right)-V|\psi|^2,
    \end{equation}
    with a potential energy $V=V(t_1,x_1,\ldots,t_N,x_N)$ discussed later (see, e.g., Refs. \cite{SE04,OIERSA19}), the Minkowski metric $g=\mathrm{diag}(1,-1)$, and the mass $m_k$.
    Note that we set $\hbar=1=c$, focus on one spatial dimension ($\alpha_k,\beta_k=0_k$ for time and $\alpha_k,\beta_k=1_k$ for space), and use Einstein's sum convention.

    For the Lagrangian density in Eq. \eqref{eq:ExtKG}, the unconstrained Euler--Lagrange equation reads
    \begin{equation}
        \label{eq:KGunconstrained}
    	0=
    	\sum_{k=1}^N\left(
            g^{\alpha_k\beta_k}
            \partial_{\alpha_k}\partial_{\beta_k}
            +m_k^2
        \right)\psi
    	+V\psi,
    \end{equation}
    where $g^{\alpha_k\beta_k} \partial_{\alpha_k}\partial_{\beta_k}=\partial_{t_k}^2-\partial_{x_k}^2$ is the d'Alembertian (aka wave operator) for the $k$th particle.
    Evaluating Eq. \eqref{eq:ReFEL2} for the factorizable wave function $\psi=\psi_\mathrm{fac}=\psi_1\cdots\psi_N$ for the purpose of obtaining the restricted equation of motion \eqref{eq:ReFEL2}, we find
    \begin{equation}
    \begin{aligned}
    	0={}&
    	\left[
            \int\prod_{l:l\neq j}
            \mathrm{d}t_l \mathrm{d}x_l\,|\psi_l|^2
    	\right]
    	\left(
            g^{\alpha_j\beta_j}
            \partial_{\alpha_j}\partial_{\beta_j}
            +m_j^2
        \right)\psi_j
    	\\
    	{}&
    	+\left[
        \sum_{k:k\neq j}
    	\left[
            \int\prod_{l:l\neq j,k}
            \mathrm{d}t_l \mathrm{d}x_l\,|\psi_l|^2
    	\right]
        \right.
    	\\
    	{}&\phantom{ + }
        \left.
        \times
        \int \mathrm{d}t_k\mathrm{d}x_k\,\psi_k^\ast
        \left(
            g^{\alpha_k\beta_k}
            \partial_{\alpha_k}\partial_{\beta_k}
            +m_k^2
        \right)\psi_k
        \right]\psi_j
    	\\
    	{}&
        + \left[
            \int\prod_{l:l\neq j}
            \mathrm{d}t_l\mathrm{d}x_l\,|\psi_l|^2
            V
        \right]
    	\psi_j
    \end{aligned}
    \end{equation}
    for the $j$th particle, $j\in\{1,\ldots,N\}$.
    Using spacetime-like expectation values,
    \begin{equation}
        \label{eq:PartiallyReducedExpectation}
    	\langle\langle \hat f\rangle\rangle_k
    	\stackrel{\text{def.}}{=}\frac{
            \int \mathrm{d}t_k\mathrm{d}x_k\,
            \psi_k^\ast \hat f\psi_k
        }{
            \int \mathrm{d}t_k\mathrm{d}x_k\,
            \psi_k^\ast\psi_k
        },
    \end{equation}
    the above equation readily simplifies to
    \begin{equation}
        \label{eq:KGconstrained}
    \begin{aligned}
        0={}&
        \left(
            g^{\alpha_j\beta_j}
            \partial_{\alpha_j}\partial_{\beta_j}
            +m_j^2
        \right)\psi_j
        \\
        {}&+\langle\langle V\rangle \rangle_{1,\ldots,j-1,j+1,\ldots N}\,\psi_j
        \\
        {}&
        -\sum_{k:k\neq j}\left(
            g^{\alpha_k\beta_k}
            \langle\langle \hat p_{\alpha_k}\hat p_{\beta_k}\rangle\rangle_k
            -m_k^2
        \right)\psi_j,
    \end{aligned}
    \end{equation}
    where $\hat p_{0_k}=\hat E_k=i\partial_{t_k}$ and $\hat p_{1_k}=\hat p_k=-i\partial_{x_k}$ denote the energy and momentum operators, respectively.

    Equations \eqref{eq:KGunconstrained} and \eqref{eq:KGconstrained} for the entangling and factorizable case, respectively, yield the distinct spacetime behavior of the two cases.
    Furthermore, we consider a second-order Taylor expansion that is akin to a spacetime-symmetric version of Hooke's law,
    \begin{equation}
        \label{eq:QuadraticPotential}
    	V=-\sum_{k,l}\Omega_{k,l}g_{\alpha_k\beta_l}x^{\alpha_k}x^{\beta_l}
    	=
    	-t^\mathrm{T}\Omega t+x^\mathrm{T}\Omega x,
    \end{equation}
    where the contravariant vector element $x^{0_j}$ corresponds to the $j$th particle's time coordinate $t_j$, i.e., $x^{0_j}=t_j$ and $t=(t_1,\ldots,t_N)$. Likewise $x^{1_j}$ corresponds to particle $j$'s spatial coordinate $x_j$ so that $x^{1_j}=x_j$ and $x=(x_1,\ldots,x_N)$.
    And the matrix $\Omega=\Omega^\mathrm{T}\in\mathbb R^{N\times N}$ describes the interaction between the $j$th particle (row) and $k$th particle (column).
    First-order terms in the Taylor series can be ignored by introducing local displacements, i.e., centralized coordinates $\Delta q_k=q_k-\langle\langle q_k\rangle\rangle_k$ for $q\in\{t,x\}$ and $k\in\{1,\ldots,N\}$.
    As an example, spacetime-distance-based interactions, mediated by a coupling constant $\kappa$, and an external potential given by $\eta$ can be introduced via the concrete potential
    \begin{equation}
    \begin{aligned}
            V
            ={}&
            -\frac{\kappa}{2} \sum_{j=1}^N \sum_{k=1}^N \left( \left(t_j - t_k\right)^2 - \left(x_j - x_k\right)^2 \right)
            \\
            {}&- \eta \sum_{j=1}^N \left(t_j^2 - x_j^2 \right)
            .
    \end{aligned}
    \end{equation}
    This corresponds to $
        \Omega
    	=
    	\left(
            \eta+N\kappa
        \right)
    	\mathbbm 1
    	-\kappa
    	\left[\begin{smallmatrix}
            1 & \hdots & 1
            \\
            \vdots & \ddots & \vdots
            \\
            1 & \hdots & 1
    	\end{smallmatrix}\right]
    $ in Eq. \eqref{eq:QuadraticPotential}, with $\mathbbm 1$ denoting the $N\times N$ identity matrix.

    Note that the kinetic and potential expressions that define the Lagrangian density under study are expressed in terms of scalar quantities, such as $g_{\alpha\beta}A^\alpha B^\beta=g^{\alpha\beta}A_\alpha B_\beta$, ensuring a proper covariant formulation of the model.

\subsection{Gaussian solutions}
\label{sec:GaussianSolutions}

    In this subsection, we restrict ourselves to analytically tractable Gaussian solutions that allow the application of continuous-variable entanglement witnesses introduced below.
    It explicitly does not characterize the general solutions.
    The full solutions, including generally non-Gaussian wave functions expressed in terms of parabolic cylinder functions \cite{MathFunctLibrary} and solutions for arbitrary subsystem partitions, are provided in Appendix \ref{app:A}.
    Moreover, we here consider massless particles, $m_1=\cdots=m_N=0$, unlike in Appendix \ref{app:A}.

    Under the above premises, including centralized coordinates, Eq. \eqref{eq:KGunconstrained} for the entangling case can be approached by using the spacetime-symmetric ansatz
    \begin{equation}
    	\psi\propto\exp\left(
            -\frac{1}{2}
            \begin{bmatrix}
            	t \\ x
            \end{bmatrix}^\mathrm{T}
            \begin{bmatrix}
            	M & C \\ C & M
            \end{bmatrix}
            \begin{bmatrix}
            	t \\ x
            \end{bmatrix}
    	\right),
    \end{equation}
    where $M=M^T\in\mathbb C^{N\times N}$ and $C=C^T\in\mathbb C^{N\times N}$.
    Substituting the wave function into Eq. \eqref{eq:KGunconstrained} for the potential in Eq. \eqref{eq:QuadraticPotential} yields
    \begin{equation}
    \begin{aligned}
    	0={}&
    	\left[
            t^\mathrm{T}(M^2-C^2-\Omega) t
            +x^\mathrm{T}(C^2-M^2+\Omega) x
    	\right.
    	\\
    	{}&
    	\left.
            +2t^\mathrm{T}(MC-CM) x
    	\right]\psi.
    \end{aligned}
    \end{equation}
    Each distinct quadratic term in this equation ought to be zero, i.e., $M^2=\Omega+C^2$ and $MC=CM$.
    Since $C$ commutes with $M$ and $M$ is a function of $\Omega$, we take
    \begin{equation}
        \label{eq:GaussMatrixMunrestricted}
    	M=(\Omega+C^2)^{1/2},
    	\quad\text{for }
    	C\text{ such that }
    	C\Omega=\Omega C
    \end{equation}
    and with the matrix square root that takes the branch with a positive real part for all eigenvalues of $M$ to ensure integrability, including a vanishing wave function for the boundary at infinity.

    Similarly, we solve the factorizable case, employing the potential in Eq. \eqref{eq:QuadraticPotential} with the matrix $\Omega$ and its entries $\Omega_{k,l}$.
    That is, we take
    \begin{equation}
    	\psi_j\propto\exp\left(
            -\frac{1}{2}
            \begin{bmatrix}
            	t_j \\ x_j
            \end{bmatrix}^\mathrm{T}
            \begin{bmatrix}
            	\mu_j & \gamma_j \\ \gamma_j & \mu_j
            \end{bmatrix}
            \begin{bmatrix}
            	t_j \\ x_j
            \end{bmatrix}
    	\right),
    \end{equation}
    for $j\in\{1,\ldots,N\}$ and $\mu_j,\gamma_j\in\mathbb C$.
    This ansatz results in the following form of Eq. \eqref{eq:KGconstrained}:
    \begin{equation}
        \label{eq:KGfactorizableIntermeadiate}
    \begin{aligned}
    	0={}&
    	\left(
            \partial_{t_j}^2
            -\partial_{x_j}^2
            -\Omega_{j,j}t_j^2
            +\Omega_{j,j}x_j^2
    	\right)\psi_j
    	\\
    	{}&
    	+\sum_{k:k\neq j}\Big(
            -\langle\langle \hat E_k^2\rangle\rangle_k
            +\langle\langle \hat p_k^2\rangle\rangle_k
    	\\
    	{}&
            -\Omega_{k,k}\langle\langle t_k^2\rangle\rangle_k
            +\Omega_{k,k}\langle\langle x_k^2\rangle\rangle_k
    	\Big)\psi_j,
    \end{aligned}
    \end{equation}
    keeping in mind that $m_k=0$, $\langle \langle t_k\rangle\rangle_k=0=\langle \langle x_k\rangle\rangle_k$ for all $k$.
    Similarly to the previous case, the second-order terms from the potential and the derivatives that are proportional to $t_j^2$, $x_j^2$, and $x_jt_j$ give
    \begin{equation}
    	\mu_j=\sqrt{\Omega_{j,j}-\gamma_j^2}
    \end{equation}
    for $j\in\{1,\ldots,N\}$ and the root's branch with nonnegative real part for integrability.
    Common Gaussian integrals also yield $
        \langle\langle x_k^2\rangle\rangle_k
        =\langle\langle t_k^2\rangle\rangle_k
    $ and $
        \langle\langle \hat E_k^2\rangle\rangle_k
        =\langle\langle \hat p_k^2\rangle\rangle_k
    $, implying that the terms in Eq. \eqref{eq:KGfactorizableIntermeadiate}, which are constant with respect to the degrees of freedom of $k$, sum to zero.
    Combining the solutions for all $j$, we can write
    \begin{equation}
    	\psi_\mathrm{fac}\propto\exp\left(
            -\frac{1}{2}
            \begin{bmatrix}
            	t \\ x
            \end{bmatrix}^\mathrm{T}
            \begin{bmatrix}
            	M_\mathrm{fac} & C_\mathrm{fac} \\ C_\mathrm{fac} & M_\mathrm{fac}
            \end{bmatrix}
            \begin{bmatrix}
            	t \\ x
            \end{bmatrix}
     	\right),
    \end{equation}
    where
    \begin{equation}
        \label{eq:GaussMatrixMrestricted}
    \begin{aligned}
        C_\mathrm{fac}={}&
        \mathrm{diag}\left(
            \gamma_1,\ldots,\gamma_N
    	\right)
    	\quad\text{and}
    	\\
    	M_\mathrm{fac}={}&\mathrm{diag}\left(
            \sqrt{\Omega_{1,1}+\gamma_1^2},
            \ldots,
            \sqrt{\Omega_{N,N}+\gamma_N^2}
    	\right).
    \end{aligned}
    \end{equation}
    The factorizable multi-time wave function only includes matrices with vanishing off-diagonal elements, i.e., spatial ($x_jx_k$), temporal ($t_jt_k$), and spatio-temporal ($x_jt_k$ and $t_jx_k$) degrees of freedom.
    Thus, the state does not include quantum correlations between the $j$th and $k$th particle for arbitrary $k\neq j$ in this relativistic scenario of $N$ interacting particles.

    Again, more general, non-Gaussian solutions for both the entangling and factorizable case are provided in Appendix \ref{app:A}.
    Furthermore, note that we can also write our solutions as
    \begin{equation}
        \label{eq:spatialRepresentation}
    \begin{aligned}
    	\psi\propto{}&
    	\exp\left(
            -\frac{1}{2}
            \begin{bmatrix}
            	t \\ x
            \end{bmatrix}^\mathrm{T}
            \begin{bmatrix}
            	M & C \\ C & M
            \end{bmatrix}
            \begin{bmatrix}
            	t \\ x
            \end{bmatrix}
    	\right)
    	\\
    	\propto{}&
    	\exp\left(
            -\frac{1}{2}\left(
                x+M^{-1}C t
            \right)^\mathrm{T} M\left(
                x+M^{-1}C t
            \right)
    	\right)
    	\\
    	{}&\times
    	\exp\left(
            -\frac{1}{2}t^\mathrm{T}\left[
                M-CM^{-1}C
            \right] t
    	\right)
    	,
    \end{aligned}
    \end{equation}
    including the factorizable case by setting $M=M_\mathrm{fac}$ and $C=C_\mathrm{fac}$ for $\psi=\psi_\mathrm{fac}$.
    This allows one to consider the spatial components as a function of times $t=(t_1,\ldots,t_N)$ and a time-dependent factor.

\subsection{Continuous-variable entanglement tests}

    To probe the entanglement for the states under study, we employ the criteria put forward in Ref. \cite{SV13} which has also been applied in experiments \cite{GSVCRTF15,GSVCRTF16}.
    Specifically, for a Gaussian state $\psi\propto\exp(-q^\mathrm{T}Qq/2)$, where $Q=Q^\mathrm{T}\in\mathbb C^{(N_1+N_2)\times (N_1+N_2)}$, we can define operators
    \begin{equation}
    \begin{aligned}
        \hat f=\nabla_q+Qq,
        \quad
        \hat f^\dag={}&-\nabla_q^\mathrm{T}+q^\mathrm{T}Q^\dag,
        \\
        \text{and}\quad
        \hat L={}&\hat f^\dag\hat f.
    \end{aligned}
    \end{equation}
    It is not hard to see that $\psi$ is the ground state of $\hat L$ because of $\hat f\psi=0$.
    This implies that the expectation value of our witness $\hat L$ for a Gaussian state $\psi$ is identically zero by construction, i.e., $\langle\langle\hat L\rangle\rangle_{\psi}=0$.

    Furthermore, we apply the optimization carried out in Ref. \cite{SV13}, in which one minimizes the expectation value of $\hat{L}$ over all non-entangled states.
    This yields the state $\psi_\mathrm{fac,min}\propto\exp(-q_1^\mathrm{T}Q_1q_1/2)\exp(-q_2^\mathrm{T}Q_2q_2/2)$ with respect to a bipartition of the degrees of freedom as $q=(q_1,q_2) \in\mathbb R^{N_1}\times\mathbb R^{N_2}$.
    More specifically, we have the block-matrix representations
    \begin{equation}
    	Q=
    	\begin{bmatrix}
            Q_{1,1} & Q_{1,2}
            \\
            Q_{1,2}^\mathrm{T} & Q_{2,2}
    	\end{bmatrix}
    	\quad\text{and}\quad
    	Q_\mathrm{fac,min}=
    	\begin{bmatrix}
            Q_{1} & 0
            \\
            0 & Q_{2}
    	\end{bmatrix},
    \end{equation}
    where
    \begin{equation}
    \begin{aligned}
        \mathrm{Im}\,Q_{1}
        ={}&
        \mathrm{Im}\,Q_{1,1},
        \quad
        \mathrm{Im}\,Q_{2}
        =
        \mathrm{Im}\,Q_{2,2},
        \\
        \mathrm{Re}\,Q_1
        ={}&\Big(
            [\mathrm{Re}\,Q_{1,1}]^2
        \\
        {}&
            +\mathrm{Re}\,Q_{1,2}\mathrm{Re}\,Q_{1,2}^\mathrm{T}
            +\mathrm{Im}\,Q_{1,2}\mathrm{Im}\,Q_{1,2}^\mathrm{T}
        \Big)^{1/2},
        \\
        \mathrm{Re}\,Q_2
        ={}&\Big(
            [\mathrm{Re}\,Q_{2,2}]^2
        \\
        {}&
            +\mathrm{Re}\,Q_{1,2}^\mathrm{T}\mathrm{Re}\,Q_{1,2}
            +\mathrm{Im}\,Q_{1,2}^\mathrm{T}\mathrm{Im}\,Q_{1,2}
        \Big)^{1/2}.
    \end{aligned}
    \end{equation}

    With this factorizable minimum state, we can determine the expectation value of $\hat L$, establishing the lower bound for factorizable states as $
        g_\mathrm{fac,min}
        =
    	\frac{1}{2}\mathrm{Tr}\left[
            (Q_\mathrm{fac,min}-Q)
            (\mathrm{Re}\,Q_\mathrm{fac,min})^{-1}
            (Q_\mathrm{fac,min}-Q)^\dag
    	\right]
    $.
    Note that this minimum is not achieved under the assumption of a Gaussianity;
    thus the entanglement test applies to non-Gaussian scenarios, too.
    It is also worth considering why the factorizable lower bound is greater than zero, $g_\mathrm{fac,min}>0$.
    Both entangled and factorizable states have to satisfy the Robertson--Schr\"odinger uncertainty relation.
    The factorizable state has to achieve this locally in each subsystem, whereas the entangled state can exploit correlations to minimize uncertainties.
    This allows the entangled state to minimize the expectation values further than its factorizable counterpart can.
    All technical details of formalism are provided in Refs. \cite{SV13,GSVCRTF15,GSVCRTF16} and their supplemental materials.
    Here, we apply this kind of entanglement criterion in a sense where $q$ can represent arbitrary spacetime coordinates, and entanglement is certified when
    \begin{equation}
        \label{eq:EntanglementTest}
    	\langle\langle\hat L\rangle\rangle-g_\mathrm{fac,min}<0
    \end{equation}
    holds true.
    Beyond bipartitions, note that the method is also rather effective for probing multipartite entanglement in theory \cite{SV13} and experiments \cite{GSVCRTF15,GSVCRTF16}.
    As discussed in Appendix \ref{app:DiffEq}, there also exist non-minimal, non-Gaussian solutions.
    Their entanglement can still be witnessed if they violate the entanglement test in Eq. \eqref{eq:EntanglementTest}.
    Importantly, however, the chosen witness may not be optimal, meaning that violation of the test is necessary but not sufficient for detecting entanglement for non-Gaussian trajectories.
    We further note that the scenario we focus on is formulated in a flat Minkowski spacetime, which can, however, be regarded as a locally flat approximation to a generally curved spacetime.

\subsection{Comparison and discussion: trajectories of two interacting particles}

    We now explicitly study the two-particle case ($N=2$) for a potential that depends on the particles' spacetime distance, $V=-\kappa[(t_1-t_2)^2-(x_1-x_2)^2]$.
    This implies a coupling matrix
    \begin{equation}
    	\Omega=
    	\begin{bmatrix}
            \kappa & -\kappa
            \\
            -\kappa & \kappa
    	\end{bmatrix}
    	=2\kappa\, e_- e_-^\mathrm{T}+0\, e_+ e_+^\mathrm{T},
    \end{equation}
    where the two orthonormal eigenvectors $e_\pm=(1,\pm 1)/\sqrt{2}$ of $\Omega$ describe center-of-mass coordinates ($+$) and difference coordinates ($-$).

\subsubsection{Explicit solutions and initial values}

    In the entangling case, the Gaussian function $\psi$ is determined through matrix $C$ that correlates temporal and spatial degrees of freedom and commutes with $\Omega$.
    In particular, Eq. \eqref{eq:GaussMatrixMunrestricted} yields
    \begin{equation}
    	C=\gamma_- e_- e_-^\mathrm{T}+\gamma_+ e_+ e_+^\mathrm{T}
    \end{equation}
    and
    \begin{equation}
    	M=
    	\sqrt{2\kappa+\gamma_-^2} e_-e_-^\mathrm{T}
    	+\sqrt{0+\gamma_+^2} e_+e_+^\mathrm{T}.
    \end{equation}
    Analogously, Eq. \eqref{eq:GaussMatrixMrestricted} for a product of single-particle Gaussian wave function $\psi_\mathrm{fac}$ in the example studied here results in the matrices
    \begin{equation}
        C_\mathrm{fac}=
        \mathrm{diag}\left(
            \gamma_1,
            \gamma_2
    	\right)
    \end{equation}
    and
    \begin{equation}
        M_\mathrm{fac}=
        \mathrm{diag}\left(
            \sqrt{\kappa+\gamma_1^2},
            \sqrt{\kappa+\gamma_2^2}
    	\right).
    \end{equation}

    For a viable comparison, we want both scenarios to start in the same initial state, $\psi(0,x_1,0,x_2)=\psi_\mathrm{fac}(0,x_1,0,x_2)=\psi_1(0,x_1)\psi_2(0,x_2)$, which is thus factorizable.
    These initial values yield $M=M_\mathrm{fac}$, readily implying
    \begin{equation}
    	\sqrt{\gamma_+^2}
    	=\sqrt{2\kappa+\gamma_-^2}
    	=\sqrt{\kappa+\gamma_1^2}
    	=\sqrt{\kappa+\gamma_2^2}
    	\stackrel{\text{def.}}{=}\mu,
    \end{equation}
    with $\mathrm{Re}(\mu)>0$.
    In other words, the initial values yield
    \begin{equation}
        \label{eq:CommonInitialValueStates}
    \begin{aligned}
    	M=M_\mathrm{fac}={}&\mu\mathbbm 1,
    	\\
    	C={}&
    	\sigma_-\sqrt{\mu^2-2\kappa}\, e_{-}e_{-}^\mathrm{T}
        +\sigma_+\mu\, e_{+}e_{+}^\mathrm{T}
    	\\
    	\text{and}\quad
    	C_\mathrm{fac}
    	={}&
    	\mathrm{diag}\left(
            \sigma_1\sqrt{\mu^2-\kappa},
            \sigma_2\sqrt{\mu^2-\kappa}
    	\right),
    \end{aligned}
    \end{equation}
    with signs $\sigma_1,\sigma_2,\sigma_+,\sigma_-\in\{+1,-1\}$ to be chosen.

\subsubsection{Trajectory entanglement}

\begin{figure}
	\includegraphics[width=.9\columnwidth]{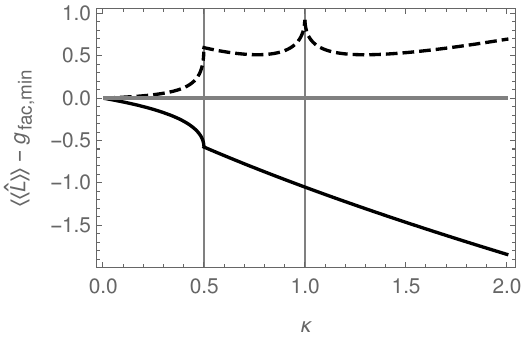}
	\caption{%
        Witnessing of entanglement [negative values, Eq. \eqref{eq:EntanglementTest}] between multi-time quantum trajectories as a function of the coupling strength $\kappa$, for $\mu=1$ and signs $\sigma_1=\sigma_2=\sigma_+=\sigma_-=1$ in Eq. \eqref{eq:CommonInitialValueStates}.
        Black, solid line shows continuous-variable entanglement of the two-particle wave function $\psi(t_1,x_1,t_2,x_2)$.
        As expected, the dashed line shows no spacetime quantum correlations for the factorizable trajectory $\psi_\mathrm{fac}(t_1,x_1,t_2,x_2)=\psi_1(t_1,x_1)\psi_2(t_2,x_2)$.
        Note that the parameters $\sqrt{\mu^2-\kappa}$ and $\sqrt{\mu^2-2\kappa}$ transition from real to imaginary values at the indicated vertical lines.
	}\label{fig:spacetime}
\end{figure}

    We apply the continuous-variable entanglement criterion in Eq. \eqref{eq:EntanglementTest} for the separation of trajectories according to $(t_1,x_1)$ and $(t_2,x_2)$.
    In Fig. \ref{fig:spacetime}, we observe that the factorizable trajectories do not lead to entanglement as required by our construction.
    By contrast, the unrestricted trajectory shows entanglement that increases (i.e., becomes more negative) with an increasing coupling strength $\kappa.$
    Also, in the weak coupling regime, $\kappa<\mu^2$, we see bifurcations effects since parameters change from real to imaginary values.
    In the strong coupling case, a smooth increase of entanglement is observable.

\begin{figure*}
	\includegraphics[height=6cm]{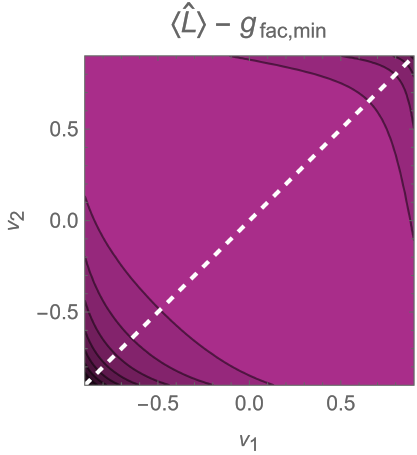}
	\quad
	\includegraphics[height=6cm]{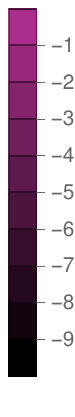}
	\hspace{1.5cm}
	\includegraphics[height=5cm]{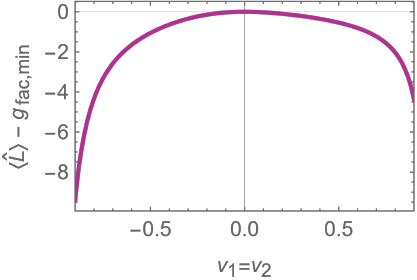}
    \\[3ex]
	\includegraphics[height=6cm]{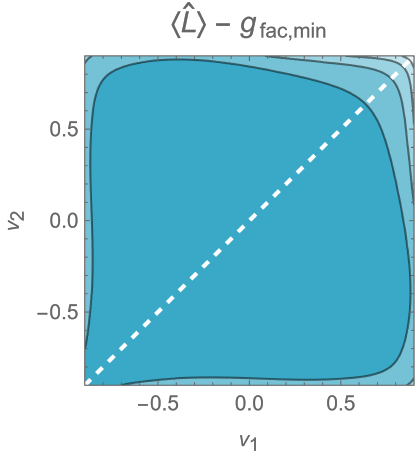}
	\quad
	\includegraphics[height=6cm]{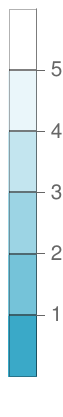}
	\hspace{1.5cm}
	\includegraphics[height=5cm]{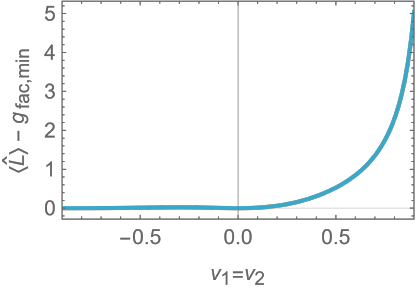}
	\caption{%
        Spatial entanglement, $\langle\hat L\rangle-g_\mathrm{fac,min}<0$, after boost operation with the particles velocities $v_1$ and $v_2$.
        Mean values are conditioned on $t_1=t_2=0$.
        Top and bottom plots show the entangled [$\psi'(0,x_1,0,x_2)$] and factorizable [$\psi_\mathrm{fac}'(0,x_1,0,x_2)$] case, respectively, both for $\mu=1$ and $\kappa=3/2$.
        Magenta-to-black shades indicate negative values and certify entanglement, with darker shades corresponding to more entanglement (negative values); cyan-to-white shades indicate nonnegative values, for which no entanglement is detected;
        the center at $v_1=0=v_2$ presents the value of zero and the initial reference frame.
        The emergence of spatial entanglement in some reference frames is a consequence of the trajectory entanglement in all reference frames.
	}\label{fig:boostedspatial}
\end{figure*}

\subsubsection{Spatial entanglement under Lorentz boosts}

    In Sec. \ref{sec:GaussianSolutions}, we already pointed out that the states in the reference frame under study are factorizable because $M=M_\mathrm{fac}$ fully characterizes the state for $(t_1,t_2)=(0,0)$, Eq. \eqref{eq:spatialRepresentation}.
    Our initial motivation was to rectify the ambiguity of other approaches to the notion of entanglement under Lorentz transformations,
    \begin{equation}
    	\begin{bmatrix}
    		t_1 \\ t_2 \\ x_1 \\ x_2
    	\end{bmatrix}
    	\mapsto
    	\begin{bmatrix}
    		\frac{1}{\sqrt{1-v_1^2}} & 0 & \frac{v_1}{\sqrt{1-v_1^2}} & 0
    		\\
    		0 & \frac{1}{\sqrt{1-v_2^2}} & 0 & \frac{v_2}{\sqrt{1-v_2^2}}
    		\\
    		\frac{v_1}{\sqrt{1-v_1^2}} & 0 & \frac{1}{\sqrt{1-v_1^2}} & 0
    		\\
    		0 & \frac{v_2}{\sqrt{1-v_2^2}} & 0 & \frac{1}{\sqrt{1-v_2^2}}
    	\end{bmatrix}
    	\begin{bmatrix}
    		t_1 \\ t_2 \\ x_1 \\ x_2
    	\end{bmatrix}.
    \end{equation}
    When transformed time coordinates are set to zero, the state is fully characterized by the matrix
    \begin{equation}
    	M'=
    	\mu\left(
            \Gamma^2
            +\Sigma^2
        \right)
        +\Gamma C\Sigma+\Sigma C\Gamma
        ,
    \end{equation}
    where
    \begin{equation}
    	\Gamma=\left[\begin{smallmatrix} \frac{1}{\sqrt{1{-}v_1^2}} & 0 \\ 0 & \frac{1}{\sqrt{1{-}v_2^2}} \end{smallmatrix}\right]
    	\quad\text{and}\quad
    	\Sigma=\left[\begin{smallmatrix} \frac{v_1}{\sqrt{1{-}v_1^2}} & 0 \\ 0 & \frac{v_2}{\sqrt{1{-}v_2^2}} \end{smallmatrix}\right]
    \end{equation}
    describe the boost's impact.
    Analogously, we can construct $M'_\mathrm{fac}$ for the factorizable scenario from $C_\mathrm{fac}$.
    Note that for fixed times, expected values are not subject to averages over time, here indicated by the common $\langle \hat L\rangle$ notation, rather than $\langle\langle\hat L\rangle\rangle$.
    It is also worth mentioning that the matrices $M'$ and $M'_\mathrm{fac}$, which characterize our general and factorizable cases, are not generally invariant under the sign reversal $v_j \mapsto -v_j$.
    While the squared quantities and the matrix $\Gamma$ are even in velocity, the matrix $\Sigma$ and the corresponding terms linear in $\Sigma$ are odd in velocity.
    Thus, one cannot expect that the entanglement of our states to be symmetric regarding the sign of the boosts, as observed in Fig. \ref{fig:boostedspatial}.

    Spatial entanglement $\psi'(0,x_1,0,x_2)$ and factorizability $\psi_1'(0,x_1)\psi_2'(0,x_2)$ as a function of distinct Lorentz transformations is shown in Fig. \ref{fig:boostedspatial}.
    The top plot shows how the observable entanglement in the spatial degrees of freedom increases (darker values) with higher absolute values of velocities of the reference frames.
    Importantly, the bottom plot demonstrates that entanglement cannot be detected for factorizable trajectories, regardless of the reference frame [all cyan shades correspond to nonnegative values for Eq. \eqref{eq:EntanglementTest}], therefore confirming the proper function of our framework in relativistic settings.
    Notably, the effects discussed become relevant primarily at relativistic speeds, i.e., for velocities approaching the speed of light, as illustrated by the right plots corresponding to equal velocities $v_1=v_2$.
    Furthermore, because of the spacetime symmetric properties of the example, we could have focused on fixed positions, e.g., $x_1=0=x_2$, and explored entanglement between only temporal degrees of freedom, $\psi(t_1,0,t_2,0)\neq\psi_1(t_1,0)\psi_2(t_2,0)$, instead.
    Finally, note that the specific quantitative features in Fig. \ref{fig:spacetime} and Fig. \ref{fig:boostedspatial} stem from the entangling dynamics of the chosen Klein--Gordon-interaction.
    These results should not be interpreted as universal fingerprints of relativistic entanglement.
    However, any interaction that generates entanglement is expected to qualitatively produce similar results in second-order expansion.
    Specifically, any interaction-induced trajectory entanglement can introduce spatial entanglement in boosted frames.
    
    Using time-resolved balanced homodyne detection, experiments can be designed to measure the predicted effects, similarly to the realizations in Refs. \cite{GSVCRTF15,GSVCRTF16}.
    This would become feasible if the precision and temporal resolution of such setups is sufficiently high to render relativistic effects measurable, analogous to high-precision atomic clock experiments effects of general relativity \cite{MGLSD18}.

%%%%%%%%%%%%%%%%%%%%%%%%%%%%%%%%%%%%%%%%%%%%%%%%%%%%%%%%%%%
%%%%%%%%%%%%%%%%%%%%%%%%%%%%%%%%%%%%%%%%%%%%%%%%%%%%%%%%%%%
%%%%%%%%%%%%%%%%%%%%%%%%%%%%%%%%%%%%%%%%%%%%%%%%%%%%%%%%%%%
\section{Conclusion}
\label{sec:Conclusion}

    In this work, we developed a methodology for studying dynamical entanglement in relativistic systems.
    Rather than imposing spatial separations only, we here used a multi-time approach to consider products of quantum trajectories that serve as uncorrelated states, and entangled trajectories that do not obey such a factorization.
    Moreover, we derived the equations of motion that govern the dynamics for the factorizable case, applying variational methods that also yield the entangling dynamics.

    Firstly, we demonstrated that in standard entanglement theory the label entangled or factorizable is not invariant under spacetime transformations that act only on individual particles.
    This means that entanglement may appear to change when viewed from different reference frames, which is not desirable for a coherent description of relativistic entanglement.
    A resolution of this issue was formulated in terms of products of trajectories, rather than products of states.
    We also showed that relaxing the underlying multi-time approach to a single time led to the known notion of single-time factorizability.
    With our approach, transforming each particle's spacetime coordinates together with their state separately does not affect factorizability, superseding the need for a common reference frame.

    Secondly, we put forward our ReFEL equations, being Euler--Lagrange equations that govern the propagation for a strictly factorizable spacetime trajectory.
    Importantly, those equations are based on the same Lagrangian that is used to determine the evolution of non-factorizable trajectories, allowing for interactions without resulting in entanglement.
    Comparing the multi-time wave functions from the restricted and unrestricted treatment then yields insight into the spacetime properties of entanglement between any number of quantum particles.
    The removal of entanglement in the ReFEL equations can also be seen as an (artificial) censorship of entanglement \cite{PS24,PM24}, allowing for a comparison with the actual (possibly) entangling dynamics.

    Finally, we applied our findings by solving the ReFEL equations for two particles obeying a Klein--Gordon-type equation with interactions proportional to their spacetime separation.
    Thereby, continuous-variable entanglement was characterized by constructing trajectory-entanglement witnesses.
    We explored the entanglement between the trajectories via a witness that jointly addresses two-particle spacetime correlations.
    Furthermore, we analyzed the impact of changing reference frames on the purely spatial entanglement via distinct Lorentz boosts for two particles.
    In the original frame, the initial values were chosen such that no purely spatial entanglement was observable, which changed after applying boost operations.
    Importantly, the factorizable trajectory did not show spatial entanglement for any relative velocities.

    Our broadly applicable framework paves the way to identify dynamical entanglement in relativistic settings.
    Possible generalizations of our approach include, for instance, extending the scalar‑valued Klein–-Gordon particles discussed here to spinor‑valued Dirac particles, and transitioning from the first‑quantized framework used in this work to a second‑quantized one, thereby allowing for entanglement between fields \cite{SA23}, because all of them can be formulated in a Lagrangian formalism.
    Moreover, a deeper analysis of factorizable quantum trajectories is fruitful for determining quantum speed limits \cite{YS24} or area laws \cite{WVHC08} which can be overcome by entangled quantum trajectories.

{To verify relativistic entanglement experimentally, spatiotemporal entanglement witnesses, as the one employed in Sec. IV. C, have to be measured.
    For instance, an interesting candidate is time-continuous homodyne detection of moving target states \cite{AM22}.
    From the analysis of such measurement data, one can infer the value of an observable using later measurement outcomes \cite{GJM13}.
    A relativistic description of such experiments would take account of the time-like separation of events from which one can infer the value of a witness in the past.
    Recently, it has been shown that such inference hints at an equivalence between spatial and temporal entanglement \cite{PM25}.

%%%%%%%%%%%%%%%%%%%%%%%%%%%%%%%%%%%%%%%%%%%%%%%%%%%%%%%%%%%
%%%%%%%%%%%%%%%%%%%%%%%%%%%%%%%%%%%%%%%%%%%%%%%%%%%%%%%%%%%
%%%%%%%%%%%%%%%%%%%%%%%%%%%%%%%%%%%%%%%%%%%%%%%%%%%%%%%%%%%

\begin{acknowledgments}
    Y.N.F. and J.S. acknowledge support of the PhoQS Grad School.
    J.P. acknowledges support from the Alexander von Humboldt Foundation (Feodor Lynen Research Fellowship).
    This work was also supported through the Ministerium f\"ur Kultur und Wissenschaft des Landes Nordrhein-Westfalen through the project PhoQC: Photonisches Quantencomputing and the \mbox{QuantERA} project QuCABOoSE.
\end{acknowledgments}

\appendix

%%%%%%%%%%%%%%%%%%%%%%%%%%%%%%%%%%%%%%%%%%%%%%%%%%%%%%%%%%%
%%%%%%%%%%%%%%%%%%%%%%%%%%%%%%%%%%%%%%%%%%%%%%%%%%%%%%%%%%%
%%%%%%%%%%%%%%%%%%%%%%%%%%%%%%%%%%%%%%%%%%%%%%%%%%%%%%%%%%%

\section{Non-Gaussian solution for various forms of multipartite entanglement}
\label{app:A}

    In this appendix, we formulate general solutions to the model of Sec. \ref{sec:Model}.
    This includes solutions for partial entanglement, e.g., arbitrary subsystem partitions, and non-Gaussian wave functions.
    Special cases of the general solutions are the Gaussian, fully factorizable cases discussed in the main part.

%%%%%%%%%%%%%%%%%%%%%%%%%%%%%%%%%%%%%%%%%%%%%%%%%%%%%%%%%%%

\subsection{Weber differential equation and parabolic cylinder functions}
\label{app:DiffEq}

    As a primer, we briefly discuss a family of differential equations and their solutions that pertain to the problems in the main text.
    Parabolic cylinder functions \cite{MathFunctLibrary}, depending on a parameter $\nu$, are the solutions of the Weber differential equation
    \begin{equation}
        0=
        \partial^2_z D_\nu (z) 
        +\left(
            \nu + \frac{1}{2} -\frac{z^2}{4}
        \right)D_\nu (z).
    \end{equation}
    In other words, solutions of the more general differential equation
    \begin{equation}
    	0=-\partial_q^2 w(q)
    	+\left( a + bq + cq^2 \right)w(q)
    \end{equation}
    with constants $a$, $b$, and $c$, can be expressed in terms of parabolic cylinder functions, viz.
    \begin{equation}
        w(q) \propto D_{\left(
            \frac{b^2}{8\sqrt{c^3}}
            -\frac{a}{2\sqrt{c}}
            -\frac{1}{2}
        \right)}\left(
            \sqrt2\sqrt[4]{c}\,q+\frac{b}{\sqrt{2}\sqrt[4]{c}}
        \right).
    \end{equation}
    Parabolic cylinder functions can be related to Hermite polynomials $H_n$,
    \begin{equation}
        D_n(z) = 2^{-n/2}e^{-\frac{z^2}{4}}H_n\left(\frac{z}{\sqrt{2}}\right),
        \quad\text{for $n\in\mathbb N$.}
    \end{equation}
    Recall that orthonormal Hermite functions read
    $
        h_n(z) = (2^n n! \sqrt{\pi})^{-1/2} e^{-z^2/2} H_n(z)
    $, where $H_n(z) = (-1)^n e^{z^2} \partial_z^n e^{-z^2}$.

%%%%%%%%%%%%%%%%%%%%%%%%%%%%%%%%%%%%%%%%%%%%%%%%%%%%%%%%%%%

\subsection{Multipartite entanglement and reduced partial differential equation}

    We begin with an $N$-partite system, which we decompose into $K$ subsystems containing $N_1,\ldots,N_K$ individual parties, where $N_1+\cdots+N_K=N$.
    Without loss of generality, we can assume that, for $q\in\{t,x\}$, we have a $K$ partition of coordinates as follows:
    \begin{equation}
    	\vec q_1=
    	\begin{bmatrix}
    		q_1 \\ \vdots \\ q_{N_1}
    	\end{bmatrix},
    	\ldots,
    	\vec q_K=
    	\begin{bmatrix}
    		q_{N_1+\cdots+N_{K-1}+1} \\ \vdots \\ q_{N_1+\cdots+N_K}
    	\end{bmatrix}.
    \end{equation}
    We now seek solutions of the form
    \begin{equation}
    	\psi(t_1,x_1,\ldots,t_N,x_N)=\psi_1(\vec t_1,\vec x_1)\cdots\psi_{K}(\vec t_K,\vec x_K)
    \end{equation}
    for studying $K$-partite entanglement, i.e., non-factorizability according to the above form.
    We considered the special cases $K=N$ of full factorizability and $K=1$ for unrestricted wave functions (likewise, $N_1=\cdots=N_K=1$ and $N_1=N$, respectively) in the main part of the paper.

    Using the $K$ partition under study, our equations of motion [mostly from Eqs. \eqref{eq:ReFEL2},  \eqref{eq:KGunconstrained}, and \eqref{eq:QuadraticPotential}] can be recast into the form
    \begin{equation}
        \label{eq:KGReFELmultipartite}
    \begin{aligned}
    	0={}&
    	\left(
            \nabla_{\vec t_j}^\mathrm{T}\nabla_{\vec t_j}
            -\nabla_{\vec x_j}^\mathrm{T}\nabla_{\vec x_j}
    	\right)\psi_j
    	\\
    	{}&
    	+
    	\left(
            -\vec t_j^\mathrm{\,T}\overleftrightarrow{\Omega}_{j,j}\vec t_j
            +\vec x_j^\mathrm{\,T}\overleftrightarrow{\Omega}_{j,j}\vec x_j
        \right)\psi_j
        \\
        {}&
        +2\left(
            \vec\tau_j^\mathrm{T}\vec t_j
            -\vec \xi_j^\mathrm{\,T}\vec x_j
        \right)\psi_j
        +m^2\psi_j
        \\
        {}&
        +\left(
            \sum_{k:k\neq j}
            \langle\langle
                \nabla_{\vec t_k}^\mathrm{T}\nabla_{\vec t_k}
                -\nabla_{\vec x_k}^\mathrm{T}\nabla_{\vec x_k}
            \rangle\rangle_{k}
        \right)\psi_j
        \\
        {}&
        -\left(
            \sum_{k,l:k,l\neq j}
            \langle\langle
                \vec t_k^\mathrm{T}\overleftrightarrow{\Omega}_{k,l}\vec t_l
                -\vec x_k^\mathrm{T}\overleftrightarrow{\Omega}_{k,l}\vec x_l
            \rangle\rangle_{k,l}
        \right)\psi_j
    	,
    \end{aligned}
    \end{equation}
    for $j\in\{1,\ldots,K\}$.
    Therein, we have a mass term $m^2=m_1^2+\cdots+m_N^2$,
    $N_k\times N_l$ submatrices $\overleftrightarrow{\Omega}_{k,l}$ of the full matrix $\Omega=\Omega^\mathrm{T}$, where $\overleftrightarrow{\Omega}_{k,l}^\mathrm{\,T}=\overleftrightarrow{\Omega}_{l,k}$, and the vectors
    \begin{equation}
        \vec \tau_j=\sum_{k:k\neq j}
        \overleftrightarrow{\Omega}_{j,k}
        \langle\langle \vec t_k\rangle\rangle_k
        \quad\text{and}\quad
        \vec \xi_j=\sum_{k:k\neq j}
        \overleftrightarrow{\Omega}_{j,k}
        \langle\langle \vec x_k\rangle\rangle_k.
    \end{equation}
    In the following, we apply a diagonalization of the submatrices, separation of variables, and parabolic cylinder functions to solve these equations.

%%%%%%%%%%%%%%%%%%%%%%%%%%%%%%%%%%%%%%%%%%%%%%%%%%%%%%%%%%%

\subsection{General solutions}

    Suppose the submatrix $\overleftrightarrow\Omega_{j,j}$ is diagonalized through orthonormal eigenvectors $\vec e_{s}$ and eigenvalues $\omega_s^2$, where $s\in\{N_1+\cdots+N_{j-1}+1,\ldots,N_1+\cdots+N_j\}\stackrel{\text{def.}}{=}\mathcal P_j$.
    Then, we can apply a separation of variables according to this diagonalization.

\paragraph{Spatial contributions.}

    We take the spatial parts for $x_s=\vec e_s^\mathrm{T}\vec x_j$ in the direction of the $s$th eigenvector.
    And we suppose
    \begin{equation}
    	\left[-\partial_{x_s}^2
    	+\omega_s^2\left(
            x_s+\frac{\vec \xi_j^\mathrm{\,T}\vec e_s}{\omega_s^2}
    	\right)^2\right]\phi_s
    	=\omega_s\left(n_s+\frac{1}{2}\right)\phi_s.
    \end{equation}
    This means, we have Hermite functions
    \begin{equation}
    \label{eq:SpatialSolutions}
    	\phi_s(x_s)=h_{n_s}\left(
            \sqrt{\omega_s}x_s+\frac{\vec \xi_j^\mathrm{\,T}\vec e_s}{\sqrt{\omega_s^3}}
    	\right).
    \end{equation}
    We can combine those solutions for all $s\in\mathcal P_j$ and find $\phi_{\mathcal P_j}=\phi_{N_1+\cdots+N_{j-1}+1}\cdots\phi_{N_1+\cdots+N_j}$ as the product of Hermite functions.
    Also, from the differential equation in Eq. \eqref{eq:KGReFELmultipartite}, we obtain the following sum over $s\in\mathcal P_j$ as
    \begin{equation}
    \label{eq:KGDiffEqPartition}
    \begin{aligned}
        {}&
    	\left(
            -\nabla_{\vec x_j}^\mathrm{T}\nabla_{\vec x_j}
            +\vec x_j^\mathrm{T}\overleftrightarrow{\Omega}_{j,j}\vec x_j
            +2\vec\xi_j^\mathrm{\,T}\vec x_j
    	\right)\phi_{\mathcal P_j}(\vec x_j)
    	\\
    	={}&
    	\left(
            \sum_{s\in\mathcal P_j}\omega_s\left[ n_s+\frac{1}{2}\right]
            -\vec \xi_j^\mathrm{\,T}
            \left[\overleftrightarrow{\Omega}_{j,j}\right]^{-1}
            \vec \xi_j
    	\right)\phi_{\mathcal P_j}(\vec x_j).
    \end{aligned}
    \end{equation}

\paragraph{Temporal contributions.}

    The previous spatial solution in Eq. \eqref{eq:SpatialSolutions} can be substituted into Eq. \eqref{eq:KGReFELmultipartite}, also using the separation $\psi_j(\vec t_j,\vec x_j)=\phi_{\mathcal P_j}(\vec x_j)T_{\mathcal P_j}(\vec t_j)$.
    This results in
    \begin{equation}
    	0=\left(
            \nabla_{\vec t_j}^\mathrm{T}\nabla_{\vec t_j}
            -\vec t_j^\mathrm{\,T}\overleftrightarrow\Omega_{j,j}\vec t_j
            -2\vec \tau_j^\mathrm{T}\vec t_j
        \right)T_{\mathcal P_j}
        +K_jT_{\mathcal P_j},
    \end{equation}
    where $K_j$ contains all terms from Eq. \eqref{eq:KGDiffEqPartition} which do not depend on the times $\vec{t_j}$.

    We can now proceed similarly to what we did for spatial degrees of freedom.
    However, we do not necessarily select Hermite functions but general parabolic cylinder functions.
    In particular, the differential equation
    \begin{equation}
    	0=\partial_{t_s}^2 T_s(t_s)
    	-\left(
            a_s
            +\left[2\vec \tau_j^\mathrm{T}\vec e_s\right] t_s
            +\left[\omega_s^2\right] t_s^2
    	\right)T_s(t_s)
    \end{equation}
    is solved via
    \begin{equation}
    	T_{s}(t_s)=D_{\nu_j}\left(
            \sqrt{2}\left[
                \sqrt{\omega_s}t_s
                +\frac{\vec \tau_j^\mathrm{T}\vec e_s}{\sqrt{\omega_s}}
            \right]
    	\right),
    \end{equation}
    where
    \begin{equation}
    	\nu_j
    	=
    	\frac{\left(\vec \tau_j^\mathrm{T}\vec e_s\right)^2}{2\omega_s^3}
    	-\frac{a_j}{2\omega_s}-\frac{1}{2}.
    \end{equation}
    From that, we obtain $T_{\mathcal P_j}$ as the product of all $T_s$ for $s\in\mathcal P_j$.
    Furthermore, the analogous sum as carried out in the spatial case yields
    \begin{equation}
    	K_j=a_{N_1+\cdots+N_{j-1}+1}+\cdots+a_{N_1+\cdots+N_j},
    \end{equation}
    for $k\in\{1,\ldots,K\}$, allowing us to select the $a$ value.

    Combining the spatial and temporal solutions yields the general solution of the partially factorizable evolution at hand.
    In  Sec. \ref{sec:KleinGordon}, parameters were chosen such that Gaussian functions sufficed for the analysis.

%%%%%%%%%%%%%%%%%%%%%%%%%%%%%%%%%%%%%%%%%%%%%%%%%%%%%%%%%%%
%%%%%%%%%%%%%%%%%%%%%%%%%%%%%%%%%%%%%%%%%%%%%%%%%%%%%%%%%%%
%%%%%%%%%%%%%%%%%%%%%%%%%%%%%%%%%%%%%%%%%%%%%%%%%%%%%%%%%%%

\end{document}